\documentclass[letterpaper,11pt]{article}

\usepackage{tikz}
\usetikzlibrary{shapes,calc,arrows,patterns,decorations.pathmorphing
	,decorations.markings}
      \usetikzlibrary{arrows.meta}

\usepackage[english]{babel}
\usepackage{theorem}
\theoremheaderfont{\itshape\bfseries}
{\theorembodyfont{\itshape}
	\newtheorem{assumption}{\textbf{Assumption}}
	\newtheorem{lemma}{\textbf{Lemma}}
	\newtheorem{definition}{\textbf{Definition}}
	\newtheorem{theorem}{\textbf{Theorem}}
	
	\newtheorem{remark}{\textbf{Remark}}
	
	\newtheorem{problem}{\textbf{Problem}}
}

\usepackage{cite}
\usepackage{amssymb}
\usepackage{algorithmic}
\usepackage{newtxtext,newtxmath}
\usepackage{amsmath}
\usepackage{amsfonts}
\usepackage{mathrsfs}
\usepackage[ruled,vlined]{algorithm2e}
\usepackage[most]{tcolorbox}

\newcommand{\T}{^{\mbox{\tiny T}}}
\newcommand{\R}{\mathbb{R}}
\newcommand{\C}{\mathbb{C}}

\newcommand{\eps}{\varepsilon}
\let\leq\leqslant
\let\geq\geqslant

\newenvironment{proof}[1][Proof]%
{\par\noindent\textit{#1:\ }}%
{\hspace*{\fill} \rule{6pt}{6pt}}
\newenvironment{proof*}[1][Proof]%
{\par\noindent\textit{#1:\ }}{}

\DeclareMathOperator{\diag}{diag}

\DeclareMathOperator{\re}{Re}

\newenvironment{system}[1]%
{\setlength{\arraycolsep}{0.5mm}\left\{ \; \begin{array}{#1}}%
	{\end{array} \right.}
\newenvironment{system*}[1]%
{\setlength{\arraycolsep}{0.5mm} \begin{array}{#1}}%
	{\end{array}}

\begin{document}

\title{Scale-free Non-collaborative Linear Protocol Design for A Class
  of Homogeneous Multi-agent Systems} 
\author{Zhenwei Liu, Ali Saberi,
  and Anton A. Stoorvogel
  \thanks{Zhenwei Liu is with College of Information Science and
    Engineering, Northeastern University, Shenyang 110819,
    China (e-mail: liuzhenwei@ise.neu.edu.cn)} 
  \thanks{Ali Saberi is with
    School of Electrical Engineering and Computer Science, Washington
    State University, Pullman, WA 99164, USA (e-mail: saberi@wsu.edu)}
  \thanks{Anton A. Stoorvogel is with Department of Electrical
    Engineering, Mathematics and Computer Science, University of
    Twente, P.O. Box 217, Enschede, The Netherlands (e-mail:
    A.A.Stoorvogel@utwente.nl)}} 

\maketitle

\begin{abstract}
  
  In this paper, we have focused on identifying a class of continuous- and discrete-time MAS for which a \emph{scale-free non-collaborative} (i.e., scale-free fully distributed) \emph{linear} protocol design is developed. We have identified conditions on agent models that enable us to design scalable linear protocols. Moreover, we show that these conditions are necessary if the agents are single input and single output. We also provide a complete design of scalable protocols for this class.
\end{abstract}


\section{Introduction}
\label{sec:introduction}

In recent decades, the synchronization problem for multi-agent systems
(MAS) has attracted substantial attention due to the wide potential
for applications in several areas, see for instance the books
\cite{bai-arcak-wen,
  bullobook,kocarev-book,mesbahi-egerstedt,ren-book,saberi-stoorvogel-zhang-sannuti,wu-book}
and references \cite{li-duan-chen-huang,saber-murray2,ren-beard}, etc.

In the synchronization literature, the communication between agents is based on measurements of the difference between the output of a specific agent and the output of neighboring agents:
\[
\zeta_i=\sum_{j=1}^Na_{ij}(y_i-y_j)
\]
where $y_i$ denotes the output of agent and $a_{ij}$ is constant for $i,j=1,\cdots,N$.

Non-collaborative protocol only uses the relative measurement $\zeta_i$ and achieves fully distributed protocols. Collaborative protocols have been traditionally presented in MAS literature (see the above books on MAS). On the other hand, collaborative protocols allows extra information exchange between neighbors. Typically, this additional information exchange consists of relative information about the difference between the state of the protocol of a specific agent and the state of the protocol of a neighboring agent using the same network.

Collaborative protocols which were introduced by \cite{li-duan-chen-huang} has been utilized to somewhat relax	
the solvability conditions for partial-state coupling as has been documented in the book \cite{saberi-stoorvogel-zhang-sannuti}.	
Loosely speaking by allowing the extra communication exchange, the solvability conditions for partial-state coupling reduced to solvability conditions for full-state coupling. Using
non-collaborative protocols, the	
solvability conditions for partial-state coupling place strong restrictions on either poles	
or zeros of the agent model (see \cite{stoorvogel-saberi-zhang-auto2017}) and, in contrast, these conditions are not required in	
full-state coupling. Moreover, some relaxation on network knowledge also occurs in full-state coupling. This should be apparent since protocol design for partial-state coupling	
requires a distributed observer which is not needed in full-state coupling.


On the other hand, most of the proposed protocols in the literature for synchronization
of MAS requires some knowledge of the communication network such as
bounds on the spectrum of the associated Laplacian matrix or the
number of agents. As it is pointed out in
\cite{studli-2017,tegling-bamieh-sandberg,tegling2019scalability,tegling-bamieh-sandberg-23},
these protocols suffer from \emph{scale fragility} where stability
properties are lost when the size of network increases, or when
communication network is altered, such as increases or decreases in
the size of sensing neighborhoods. 

In the past few years, a scale-free protocol design has been the subject of current research for MAS. Scale-free protocol design addresses this issue by
designing protocols which do not rely on any knowledge about the
communication graph, i.e.,
\begin{enumerate}
	\item The protocol is designed only based on knowledge of the
	agent model ($A,B,C$).
	\item The protocol is designed to work with any fixed communication
	graph which contains a spanning tree without incorporating
	knowledge about the graph into the protocol.
\end{enumerate}

Almost all results of scalable protocols available in the literature are collaborative, see \cite{chowdhury-khalil,donya-liu-saberi-stoorvogel-CCC-6,liu-saberi-stoorvogel-donya-inputsaturation-automatica,liu-saberi-stoorvogel-donya-almost-automatica}. To the best of our knowledge, the scalable non-collaborative protocols are only for MAS with passive or passifiable agents, see \cite{chopra-tac} and \cite{liu-zhang-saberi-stoorvogel-auto}.

%

In this paper, the main objective is to show when it is possible to achieve	
a scale-free design which is non-collaborative and hence only relies on the original relative	
measurement $\zeta_i$. We present necessary conditions and design protocols to	
achieve this objective under assumptions which are very close to these necessary conditions.	
More specifically we have identified one class of continuous- and discrete-time	
MAS for which scalable non-collaborative (i.e., scalable fully distributed) linear protocols can be	
designed. 

%
\subsection*{Notations and Background}
Given a matrix $A\in \mathbb{R}^{m\times n}$, $A\T$ and $A^*$ denote
its transpose and conjugate transpose respectively. A square matrix
$A$ is said to be Hurwitz stable if all its eigenvalues are in the
open left half complex plane, and $A$ is said to be Schur stable if
all its eigenvalues are in the open unit disk. $A\otimes B$ depicts
the Kronecker product between $A$ and $B$. $I_n$ denotes the
$n$-dimensional identity matrix and $0_n$ denotes $n\times n$ zero
matrix; sometimes we drop the subscript if the dimension is clear from
the context.

To describe the information flow among the agents we associate a
\emph{weighted graph} $\mathcal{G}$ to the communication network. The
weighted graph $\mathcal{G}$ is defined by a triple
$(\mathcal{V}, \mathcal{E}, \mathcal{A})$ where
$\mathcal{V}=\{1,\ldots, N\}$ is a node set, $\mathcal{E}$ is a set of
pairs of nodes indicating connections among nodes, and
$\mathcal{A}=[a_{ij}]\in \mathbb{R}^{N\times N}$ is the weighted
adjacency matrix with non negative elements $a_{ij}$. Each pair in
$\mathcal{E}$ is called an \emph{edge}, where $a_{ij}>0$ denotes an
edge $(j,i)\in \mathcal{E}$ from node $j$ to node $i$ with weight
$a_{ij}$. Moreover, $a_{ij}=0$ if there is no edge from node $j$ to
node $i$. We assume there are no self-loops, i.e.\ we have
$a_{ii}=0$. A \emph{path} from node $i_1$ to $i_k$ is a sequence of
nodes $\{i_1,\ldots, i_k\}$ such that $(i_j, i_{j+1})\in \mathcal{E}$
for $j=1,\ldots, k-1$. A \emph{directed tree} is a subgraph (subset of
nodes and edges) in which every node has exactly one parent node
except for one node, called the \emph{root}, which has no parent
node. A \emph{directed spanning tree} is a subgraph which is a
directed tree containing all the nodes of the original graph. If a
directed spanning tree exists, the root has a directed path to every
other node in the tree \cite{royle-godsil}.

For a weighted graph $\mathcal{G}$, the matrix
$L=[\ell_{ij}]$ with
\[
\ell_{ij}=
\begin{system}{cl}
	\sum_{k=1}^{N} a_{ik}, & i=j,\\
	-a_{ij}, & i\neq j,
\end{system}
\]
is called the \emph{Laplacian matrix} associated with the graph
$\mathcal{G}$. The Laplacian matrix $L$ has all its eigenvalues in the
closed right half plane and at least one eigenvalue at zero associated
with right eigenvector $\textbf{1}$ \cite{royle-godsil}. Moreover, if
the graph contains a directed spanning tree, the Laplacian matrix $L$
has a single eigenvalue at the origin and all other eigenvalues are
located in the open right-half complex plane \cite{ren-book}. 

A row stochastic matrix $D$ can be associated with a graph
$\mathcal{G}$. $N$, the dimension of $D$, is the number of node and an
arc $(j,i)\in \mathcal{E}$ if $d_{ij}>0$. It is shown in
\cite{ren-beard} that 1 is a simple eigenvalue of $D$ if and only if
$\mathcal{G}$ contains a directed spanning tree. Moreover, the other
eigenvalues are in the open unit disk if $d_{ij}>0$ for all~$i$.

\section{Problem formulation}

Consider a homogeneous MAS composed of $N$ identical linear time-invariant agents of the form,
\begin{equation}\label{homo_sys}
  \begin{system*}{cl}
    {x}_i^+&=Ax_i+Bu_i,\\
    y_i&=Cx_i,
  \end{system*}
\end{equation}
where $x_i\in\mathbb{R}^{n}$, $u_i\in\mathbb{R}^{m}$ and
$y_i\in\mathbb{R}^p$ are the state, input, output of agent $i$ for
$i=1,\ldots, N$. In the aforementioned presentation, for
continuous-time systems, $x_i^+(t) = \dot{x}_i(t)$ for
$t \in \mathbb{R}$; while for discrete-time systems,
$x_i^+(t) = x_i(t + 1)$ for $t \in \mathbb{Z}$. 

The communication network is composed of $N$ linear combinations and
each combination includes agent's own output relative to that of other
agents. Network with \textbf{continuous-time} agent is shown as follows for agent $i$:
\begin{equation}\label{zeta1}
  \zeta_i=\sum_{j=1}^{N}a_{ij}(y_i-y_j)
\end{equation}
where $a_{ij}>0$ and $a_{ii}=0$. Here we use a weighted and directed
graph $\mathcal{G}$ to describe the communication topology of the
network, the nodes of network correspond to the agents and the weight
of edges given by the coefficient $a_{ij}$. In the matter of the
coefficients of the associated Laplacian matrix
$L=[\ell_{ij}]_{N\times N}$, $\zeta_i$ can be rewritten as
\begin{equation}\label{zeta}
	\zeta_i= \sum_{j=1}^{N}\ell_{ij}y_j.
\end{equation}
We refer to \eqref{zeta} as \emph{partial-state coupling} since only
part of the states are communicated over the network. When $C=I$, we
call it \emph{full-state coupling}. 

In the case of networks with \textbf{discrete-time} agents, each agent
has access to the following information
\begin{equation}\label{zetabar2-d0} 
  \zeta_i(t)=\frac{1}{1+\bar{d}_{\text{in}}(i)}\sum_{j=1,j\neq i}^N a_{ij}(y_{i}(t)-y_{j}(t))
\end{equation}
where $\bar{d}_{\text{in}}(i)$ is an \emph{upper bound} of $d_{in}(i)=\sum_{j=1}^{N}a_{ij}$ for $i=1,\ldots,N$. Next we write $\zeta_i$ as
\begin{equation}\label{zeta-y}
  \zeta_i(t)=\sum_{j=1,j\neq i}^N d_{ij}(y_i(t)-y_j(t)),
\end{equation}
where $d_{ij}\geq 0$, and we choose
$d_{ii}=1-\sum_{j=1,j\neq i}^Nd_{ij}$ such that $\sum_{j=1}^Nd_{ij}=1$
with $i,j\in\{1,\ldots,N\}$. Note that $d_{ii}$ satisfies
$d_{ii}>0$. The weight matrix $D=[d_{ij}]$ is then a so-called, row
stochastic matrix, where all eigenvalues of $D$ satisfy $|\lambda_i|\leq 1$ and 1 has one simple eigenvalue. Let $D_{\text{in}}=\diag\{\bar{d}_{\text{in}}(i)\}$. Then the
relationship between the row stochastic matrix $D$ and the Laplacian
matrix $L$ is
\begin{equation}\label{hodt-LD}
  (I+D_{\text{in}})^{-1}L=I-D.
\end{equation}

Our goal is to achieve state synchronization, i.e.,
\begin{equation}\label{synch_state}
  \lim_{t\to\infty}(x_i(t)-x_j(t))=0
\end{equation} 
for all $i,j \in \{1,\ldots, N\}$. 

We need the following definition to explicitly state our problem
formulation.

\begin{definition}\label{def1}  
  We define the following set. $\mathbb{G}^N$ denotes the set of
  fixed, directed graphs of $N$ agents which contains a directed spanning
  tree.
\end{definition}

We formulate the scale-free state synchronization problem of a MAS
without localized collaborative information exchange,
i.e. non-collaborative protocol, as follows.

\begin{problem}\label{prob4}
  The \textbf{scale-free state synchronization problem without
    localized collaborative information exchange} for MAS
  \eqref{homo_sys} and communication network with \eqref{zeta} for
  continuous-time case or \eqref{zeta-y} for discrete-time case is to
  find, if possible, a fixed linear protocol of the form:
  \begin{equation}\label{protoco1}
    \begin{system}{cl}
      x_{i,c}^+&=A_{c}x_{i,c}+B_{c}{\zeta}_i,\\
      u_i&=F_{c}x_{i,c}+G_{c}{\zeta}_i,
    \end{system}
  \end{equation}
  where  $x_{c,i}(t)\in\R^{n_c}$ is the state of protocol, and matrices $A_{c}, B_{c}, F_{c}, G_{c}$ are pre-designed parameters, such that the state synchronization \eqref{synch_state} is achieved for any number of
  agents $N$, any fixed communication graph $\mathcal{G}$ and all initial
  conditions of agents.
\end{problem}

\begin{remark}
  Note that the number of agents $N$ and the weight $a_{ij}$ are fixed in a control period.
	
  In discrete-time MAS, we work with a row stochastic matrix
  which is a scaled version of the Laplacian matrix. This is
  completely in line with all papers in this area. For the scaling it
  should be noted that this only uses some local information about the
  graph, namely, $\bar{d}_{\text{in}}(i)$ as used in \eqref{zetabar2-d0}. One might
  ask whether scale-free state synchronization problem without
  localized collaborative information exchange is possible without
  using this scaled Laplacian. It can actually be shown that this
  latter problem is never solvable in discrete-time MAS.
\end{remark}

\section{Necessary conditions for solvability}

The first important result that we provides necessary conditions
for the solvability of Problem \ref{prob4} for both continuous- and discrete-time MAS. 

\begin{theorem}[Continuous-time MAS]\label{theo_nec1}
	Consider a continuous-time, single-input, single-output MAS
	\eqref{homo_sys} with communication via \eqref{zeta}. There exists a
	linear protocol of the form \eqref{protoco1} which achieves
	scale-free state synchronization problem without localized
	collaborative information exchange only if:
	\begin{enumerate}
		\item Agent model is stabilizable and detectable.
		\item Agent model is neutrally stable. 
		\item Agent model is weakly minimum phase.
		\item Agent model has relative degree equal to $1$.
	\end{enumerate}
\end{theorem}

\begin{proof}
	The necessity of stabilizability and detectability is obvious. If we
	define
	\begin{equation}\label{ABCtilde}
		\tilde{A}=\begin{pmatrix} A & BF_c \\ 0 & A_c \end{pmatrix},\qquad
		\tilde{B}=\begin{pmatrix} BG_c \\ B_c \end{pmatrix},\qquad
		\tilde{C}=\begin{pmatrix} C & 0 \end{pmatrix}
	\end{equation}
	then \cite[Chapter 2]{saberi-stoorvogel-zhang-sannuti} has shown
	that we achieve synchronization if
	\[
	\tilde{A}+\lambda_i \tilde{B}\tilde{C}
	\]
	is asymptotically stable for all nonzero eigenvalues
	$\{\lambda_2,\ldots,\lambda_N\}$ of the Laplacian matrix $L$. To
	obtain a scale-free design we should therefore guarantee that
	\begin{equation}\label{abc}
		\tilde{A}+\lambda_i \tilde{B}\tilde{C}
	\end{equation}
	is asymptotically stable for all $\lambda_i\in \C$ with
	$\re (\lambda_i) > 0$, $i=2,\cdots,N$. We define
	\[
	g(s)=C(sI-A)^{-1}B,\qquad
	g_c(s)=F_c(sI-A_c)^{-1}B_c.
	\]
	Note that \eqref{abc} asymptotically (Hurwitz) stable is equivalent to:
	\[
	1-\lambda_i g(s)g_c(s) \neq 0
	\]
	for all $s\in \C$ with $\re s\geq 0$ without unstable pole-zero
	cancellations in $g(s)g_c(s)$. Since this must be true for all
	$\lambda_i\in \C$ with $\re (\lambda_i) >0$, this yields the requirement
	that $g(s)g_c(s)$ is positive-real. From \cite[Section
	3.51]{ioannou-sun} this requires that $g(s)g_c(s)$ satisfies:
	\begin{itemize}
		\item The poles of $g(s)g_c(s)$ are in the closed left half plane
		and the poles on the imaginary axis are simple.
		\item The zeros of $g(s)g_c(s)$ are in the closed left half plane
		and the zeros on the imaginary axis are simple.
		\item The relative degree of $g(s)g_c(s)$ is less than or equal to
		$1$.
	\end{itemize}
	Since there are no unstable pole-zero cancellations in $g(s)g_c(s)$,
	the above conditions immediately yield that the agent model should be neutrally stable, weakly minimum-phase, and have relative degree~$1$.
\end{proof}
\begin{theorem}[Discrete-time MAS]\label{theo_nec2}
	Consider a discrete-time, single-input, single-output MAS
	\eqref{homo_sys} with communication via \eqref{zeta-y}. There exists a
	linear protocol of the form \eqref{protoco1} which achieves scale-free state synchronization problem without
	localized collaborative information exchange only if:
	\begin{enumerate}
		\item Agent model is stabilizable and detectable.
		\item Agent model is neutrally stable. 
	\end{enumerate}
\end{theorem}

\begin{proof}
	The necessity of stabilizability and detectability is obvious, too. Using
	\eqref{ABCtilde}. we obtain from \cite[Chapter
	3]{saberi-stoorvogel-zhang-sannuti} that we need
	\begin{equation}
		\tilde{A}+(1-\lambda_i) \tilde{B}\tilde{C}
	\end{equation}
	is asymptotically (Schur) stable for all $\lambda_i\in \C$ with
	$|\lambda_i|<1$. Using similar arguments as in the continuous time, we
	obtain that we need that $g(s)g_c(s)+\tfrac{1}{2}$ has to be
	positive real. From \cite{xiao-djhill} we obtain that this requires
	that the poles of $g(s)g_c(s)$ are in the closed unit disc and the
	poles on the unit circle are simple. Since there are no unstable
	pole-zero cancellations in $g(s)g_c(s)$, this
	immediately yields that the agent model should be neutrally stable.
\end{proof}

\section{Scale-free non-collaborative protocol design: Continuous-time case}

We make the following assumption for agent models.

\begin{assumption}\label{ass1}
  Continuous-time agents \eqref{homo_sys} satisfy the following properties:
  \begin{enumerate}
  \item Agent model is stabilizable and detectable.
  \item Agent model is neutrally stable. 
  \item Agent model is minimum phase.
  \item Agent model must be uniform rank with order of infinite zero equal to one.
  \end{enumerate}
\end{assumption}

\begin{remark}
  If we compare the above with the necessary conditions we obtained
  for SISO systems in Theorem \ref{theo_nec1} then we note that we only
  strengthened to condition of weakly minimum-phase to
  minimum-phase. The other conditions are the same.

  We would like to emphasize that the agent model can be non-square
  and neither right nor left invertible. Also it is known that passive
  agents satisfy these Assumptions \ref{ass1} and as such form a
  subset of the class of agents that we consider in this paper.
\end{remark}
\begin{figure}[h]
	\includegraphics[width=8cm]{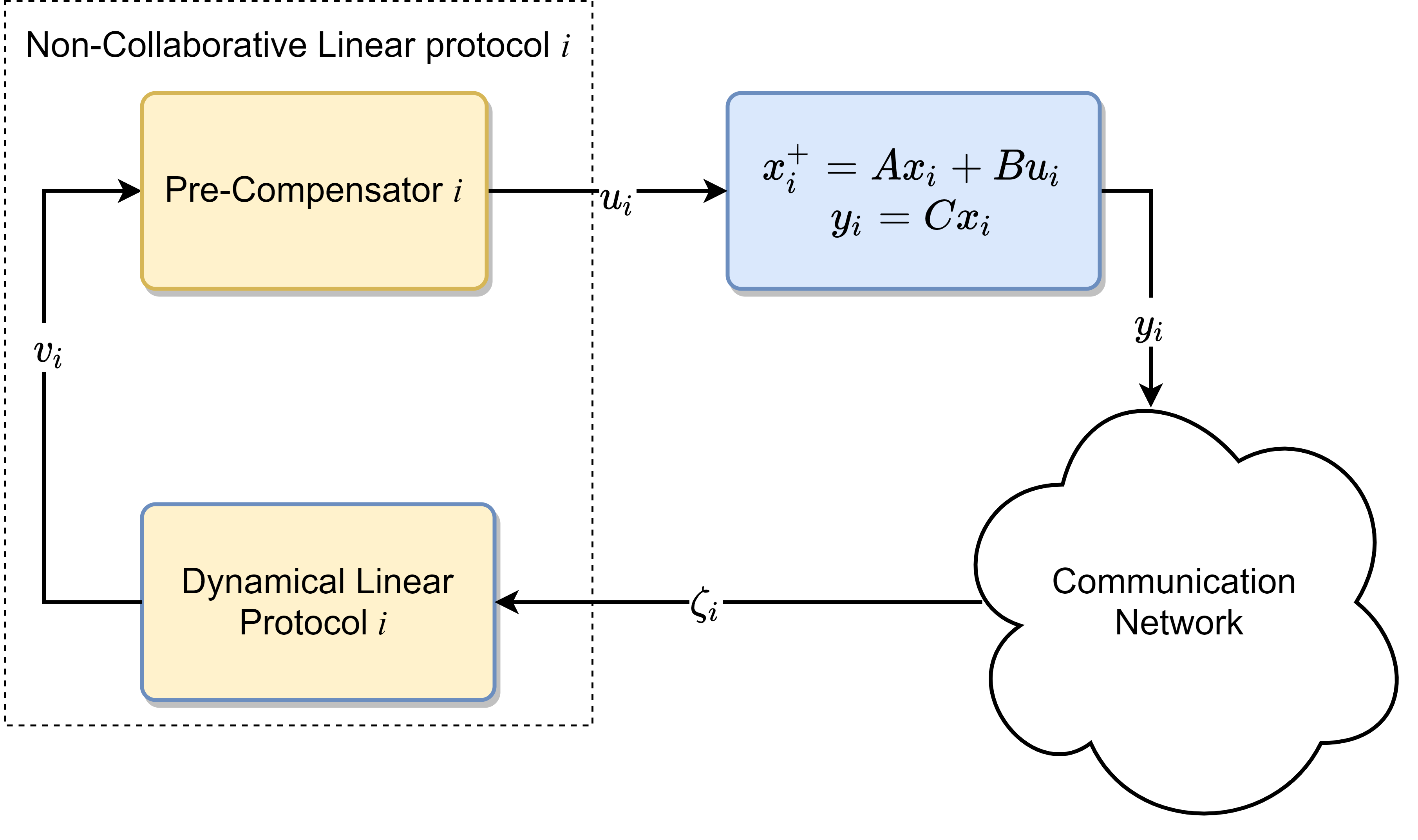}
	\centering
	\caption{Architecture of the scalable non-collaborative linear protocol}\label{noncollaborative}
\end{figure}

We provide a scale-free non-collaborative linear protocol design in
continuous via partial-state coupling. The design architecture is
shown in Fig. \ref{noncollaborative}.
In other words, the design has two steps:
\begin{enumerate}
\item The first module designs a precompensator to make the agent
  model \eqref{homo_sys} left-invertible.
\item The second module designs a non-collaborate dynamical
  protocols for left-invertible agents to achieve state
  synchronization.
\end{enumerate}

\subsection{Protocol design for partial-state coupling}\label{sub1}


The detailed design procedure is listed as follows.\medskip

\noindent\textbf{Step I: Design of pre-compensator}


\begin{figure}[h]
	\centering
	\begin{tikzpicture}[every node/.style={outer sep=0pt,thick},
		decorate, scale=0.25]
		\draw (-2,0) rectangle (8,4) node[pos=.5] {\small Pre-Compensator};
		\draw (12,0) rectangle (20,4) node[pos=.5] {\small Agent Model};
		\draw[-{Latex[length=2mm]}]  (-6,2) -- (-2,2);
		\draw[-{Latex[length=2mm]}]  (8,2) -- (12,2);
		\draw[-{Latex[length=2mm]}]  (20,2) -- (24,2);
		\node at (-4,2.7)  {$v_i$};
		\node at (10,2.7)  {$u_i$};
		\node at (22,2.8)  {$y_i$};
	\end{tikzpicture}
	\caption{The compensated agent with pre-compensator}\label{gpass2}
\end{figure}
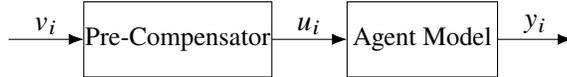 

In this step we design the following asymptotically stable
pre-compensator such that the compensated agent shown at Fig. \ref{gpass2} is left-invertible and satisfies Assumption
\ref{ass1}.
\begin{tcolorbox}[breakable,colback=white]
  \begin{equation}\label{precomz}
    \begin{system}{rl}
      \dot{p}_i=&A_pp_i+B_pv_i\\
      u_i=&C_pp_i+D_pv_i
    \end{system}	
  \end{equation}
  where $p_i\in \R^q$ and $v_i\in \R^{m_v}$ are state and input of pre-compensator. All eigenvalues of $A_p$
  are in open left-half plane.
\end{tcolorbox}

The following lemma guarantees the existence of this pre-compensator.
\begin{lemma}\label{preclem}
  Consider a continuous-time agent of the form \eqref{homo_sys} which is stabilizable
  and detectable. In that case there exists an asymptotically stable
  pre-compensator \eqref{preclem}, such that the interconnection of
  \eqref{homo_sys} and this pre-compensator which is given by,
  \begin{equation}\label{newsys}
    \begin{system}{rl}
      \dot{z}_i&=\tilde{A}{z}_i+\tilde{B}v_i\\
      y_i&=\tilde{C}z_i
    \end{system}
  \end{equation}
  where
  \[
    z_i=\begin{pmatrix}
      x_i\\
      p_i
    \end{pmatrix}, \tilde{A}=\begin{pmatrix}
      A&BC_p\\0&A_p
    \end{pmatrix}, \tilde{B}=\begin{pmatrix}
      BD_p\\B_p
    \end{pmatrix}, 
    \tilde{C}=\begin{pmatrix}
      C&0
    \end{pmatrix}.
  \]
  has the following properties:
  \begin{itemize}
  \item It is stabilizable and detectable,
  \item It is left-invertible,
  \item Its poles are the poles of the system \eqref{homo_sys} plus
    the stable poles of the pre-compensator (i.e., the eigenvalues of $A_p$),
  \item Its infinite zero structure is the same as the infinite zero
    structure of the system \eqref{homo_sys},
  \item Its invariant zeros are the invariant zeros of the system
    \eqref{homo_sys} and some additional invariant zeros that can be
    arbitrarily placed in the open left-half complex plane,
  \end{itemize}	
\end{lemma}

\begin{proof}
  Obviously, we just need to prove the case where agent model is not left-invertible, i.e., right-invertible and neither left-invertible or right-invertible.
  
  If the agent model is right-invertible, we can directly use the results in
  \cite[Section III-B, and the dual results of Theorem 3.1 and Remark 3.3]{ss} or \cite[Theorem
  1-(2) and Remark 1]{sannuti-saberi-zhang-auto}.
  
  If the agent model is neither left-invertible or right-invertible, we can design a pre-compensator only to make the compensated agent left-invertible, by using the results in
  \cite[Section III-C, and the dual results of Theorem 3.1 and Remark 3.3]{ss} or \cite[Theorem
  1-(3) and Remark 1]{sannuti-saberi-zhang-auto}.
\end{proof}\medskip

\noindent\textbf{Step II: Design of a scalable non-collaborative linear protocol}


Under Assumption \ref{ass1}, we can use the Special Coordinate Basis (SCB) \cite{ss} to achieve the
following transformation for the compensated agents
\eqref{newsys}. In other words, there exists a non-singular state transformation
matrix $S$ with
\[
\begin{pmatrix}
  \bar{z}_{1i}\\
  \bar{z}_{2i}
\end{pmatrix}=Sz_i,
\]
such that the dynamics of $\bar{z}_{1i}$ and $\bar{z}_{2i}$ are represented by
\begin{equation}\label{eq6}
  \begin{system}{cl}
    \dot{\bar{z}}_{1i}&= A_{11}\bar{z}_{1i}+A_{12}\bar{z}_{2i},\\
    \dot{\bar{z}}_{2i}&=A_{21}\bar{z}_{1i}+A_{22}\bar{z}_{2i}+\bar{B}v_i,\\
    y_i&=\begin{pmatrix}
      y_{1i}\\y_{2i}
    \end{pmatrix}=\begin{pmatrix}
      \bar{C}\bar{z}_{1i}\\\bar{z}_{2i}
    \end{pmatrix},
  \end{system}
\end{equation}
where $\bar{z}_{1i}\in\R^{n+q-\bar{n}}$ and
$\bar{z}_{2i}\in\R^{\bar{n}}$, $\bar{B}$ is a non-singular matrix, and
$(A_{11}, \bar{C})$ is detectable while
\[
  S\tilde{A}S^{-1}=\begin{pmatrix}
    A_{11}&A_{12}\\A_{21}&A_{22}
  \end{pmatrix}, S\tilde{B}=\begin{pmatrix}
    0\\\bar{B}
  \end{pmatrix}, \tilde{C}S^{-1}=\begin{pmatrix}
    \bar{C}&0\\0&I
  \end{pmatrix}.
\]
Meanwhile, we have
\[
  \zeta_{i}=\begin{pmatrix}
    \zeta_{1i}\\\zeta_{2i}
  \end{pmatrix} =\begin{pmatrix}
    \bar{C}\zeta_{si}\\\zeta_{2i}
  \end{pmatrix},\qquad
  \zeta_{si}=\sum_{j=1}^{N}\ell_{ij} \bar{z}_{1j},\qquad
  \zeta_{2i}=\sum_{j=1}^{N}\ell_{ij} \bar{z}_{2j}.
\]




Since the compensated agents are neutrally stable, we have the eigenvalues of $A$
are on the imaginary axis, if any, are semi-simple. According to Lemma
\ref{preclem}, the poles of compensated system \eqref{newsys} are on
closed left half plane and therefore, there exists a positive definite
matrix $P$ such that
\begin{equation}
  P\tilde{A}+\tilde{A}\T P\leq 0\label{codct}
\end{equation}
Now we are ready to give our scale-free protocol design below.
\begin{tcolorbox}[breakable,colback=white]
  \begin{equation}\label{ncsfpvpscct}
    \begin{system}{ccl}
      \dot{p}_i&=&A_pp_i-\rho B_p\tilde{B}\T P S^{-1}\left[\begin{pmatrix}
        I_{n+q-\bar{n}}\\0
      \end{pmatrix}\hat{\bar{z}}_{1i}+\begin{pmatrix}
      0&0\\0&I_{\bar{n}}
  \end{pmatrix}\zeta_{i}\right]\\
      \dot{\hat{\bar{z}}}_{1i}&=&(A_{11}-H\bar{C})\hat{\bar{z}}_{1i}+
      {\setlength{\arraycolsep}{0.7mm} \begin{pmatrix}
        H&A_{12}
      \end{pmatrix}}\zeta_{i}\\
      u_i&=&C_pp_i-\rho D_p\tilde{B}\T P S^{-1}\left[\begin{pmatrix}
      	I_{n+q-\bar{n}}\\0
      \end{pmatrix}\hat{\bar{z}}_{1i}+\begin{pmatrix}
      	0&0\\0&I_{\bar{n}}
      \end{pmatrix}\zeta_{i}\right]
    \end{system}
  \end{equation}
  where $H$ is a matrix such that $A_{11}-H\bar{C}$ is Hurwitz stable,
  $P>0$ satisfies \eqref{codct}, $\rho >0$, and $i=1,\ldots,N$.
\end{tcolorbox}

Next, we have the following theorem to achieve state synchronization.
\begin{theorem}\label{th-linear}
  Consider a continuous-time MAS described by \eqref{homo_sys} and
  \eqref{zeta1}. Assume Assumption \ref{ass1} is satisfied. Let the
  set $\mathbb{G}^N$ denote all graphs satisfy Definition \ref{def1}.
	
  Then, the scale-free non-collaborative state synchronization problem
  via linear protocol as stated in Problem \ref{prob4} is
  solvable. More specifically, the protocol \eqref{ncsfpvpscct}
  achieve state synchronization for any fixed graph
  $\mathscr{G} \in \mathbb{G}^N$ with any size of the network $N$.
\end{theorem}

\begin{proof}
  According to Lemma \ref{preclem}, we can know that there must exist
  pre-compensator \eqref{precomz} to make agent \eqref{homo_sys}
  left-invertible, and obtain the compensated system \eqref{newsys}.
	
  From \eqref{eq6} and \eqref{ncsfpvpscct}, we have
  \begin{align*}
    \dot{\bar{z}}_{1i}&= A_{11}\bar{z}_{1i}+A_{12}\bar{z}_{2i}\\
    \dot{\hat{\bar{z}}}_{1i}&=
    A_{11}\hat{\bar{z}}_{1i}+A_{12}\zeta_{2i}+H(\zeta_{1i}-\bar{C}\hat{\bar{z}}_{1i}).
  \end{align*}	
  By defining 
  \[
    \bar{z}_1=\begin{pmatrix}
      \bar{z}_{11}\\\vdots\\\bar{z}_{1N}
    \end{pmatrix}, \qquad\bar{z}_2=\begin{pmatrix}
      \bar{z}_{21}\\\vdots\\\bar{z}_{2N}
    \end{pmatrix}, \qquad\hat{\bar{z}}_1=\begin{pmatrix}
      \hat{\bar{z}}_{11}\\\vdots\\\hat{\bar{z}}_{1N}
    \end{pmatrix},
  \]
  we obtain
  \begin{align*}
    \dot{\bar{z}}_{1}&= (I\otimes A_{11})\bar{z}_{1}+(I\otimes A_{12})\bar{z}_{2}\\
    \dot{\hat{\bar{z}}}_{1}&= [I\otimes
    (A_{11}-H\bar{C})]\hat{\bar{z}}_{1}+(L\otimes
    A_{12})\bar{z}_{2}+(L\otimes H\bar{C})\bar{z}_{1} 
  \end{align*}
  Let $e=(L\otimes I)\bar{z}_{1}-\hat{\bar{z}}_{1}$, then we have
  \[
    \dot{e}=[I\otimes (A_{11}-H\bar{C})]e.
  \]
  Since $A_{11}-H\bar{C}$ is Hurwitz stable, it is obvious that $e$ is
  asymptotically stable, i.e.
  \[
    \lim_{t\to \infty} \hat{\bar{z}}_{1i}\to \zeta_{si}=\sum_{j=1}^{N}\ell_{ij} \bar{z}_{1j}.
  \]
  Meanwhile, we obtain 
  \begin{equation}\label{convereq}
    \begin{pmatrix}
      \hat{\bar{z}}_{1i}\\\zeta_{2i}
    \end{pmatrix}\to S\left(\sum_{j=1}^N\ell_{ij} z_j\right) \text{ as }t\to \infty.
  \end{equation}
  On the other hand, from \eqref{precomz}, \eqref{newsys}, and \eqref{ncsfpvpscct} we have
  \[
    v_i=-\rho\tilde{B}\T P S^{-1}\begin{pmatrix}
      \hat{\bar{z}}_{1i}\\\zeta_{2i}
    \end{pmatrix}
  \]
  According to agent model \eqref{newsys} and result \eqref{convereq}, we have
  \[
    \dot{z}_i=\tilde{A}z_i-\rho\tilde{B}\tilde{B}\T P\sum_{j=1}\ell_{ij} z_j
  \]
  Then, by setting
  \[
    z=\begin{pmatrix}
      z_{1}\\\vdots\\z_{N}
    \end{pmatrix}, 
  \]
  we obtain
  \begin{equation}\label{cls2}
    \dot{z}=(I\otimes \tilde{A}-\rho L\otimes \tilde{B}\tilde{B}\T P)z.
  \end{equation}
  
  By using the method from \cite[Lemma 2]{liu-zhang-saberi-stoorvogel-auto},
  there exists a non-singular matrix $T$ such that \eqref{cls2} can be
  transformed as
  \begin{equation}\label{cls3}
    \begin{system}{l}
      \dot{\eta}_1=\tilde{A}\eta_1,\\
      \dot{\eta}_i=(\tilde{A}-\rho\lambda_i \tilde{B}\tilde{B}\T P)\eta_i, \qquad i=2,\ldots, N,
    \end{system}
  \end{equation}
  where $\lambda_i$ denotes all non-zero eigenvalues of
  $L$. Therefore, we need to prove the stability of \eqref{cls3} to
  obtain original MAS' state synchronization, i.e. the stability of
  $\tilde{A}-\rho\lambda_i \tilde{B}\tilde{B}\T P$ for $i=2,\ldots, N$
  where we know that $\re(\lambda_i)>0$, i.e.\ the real part of
  $\lambda_i$ is positive.
	
  Choosing $P>0$ satisfying \eqref{codct}, then we have
  \begin{align*}
    &P(\tilde{A}-\rho\lambda_i \tilde{B}\tilde{B}\T
    P)+(\tilde{A}-\rho\lambda_i \tilde{B}\tilde{B}\T P)^*P\\ 
    =&P\tilde{A}+\tilde{A}\T P-2\rho\re(\lambda_i)P\tilde{B}\tilde{B}\T P\\
    \leq &-2\rho\re(\lambda_i)P\tilde{B}\tilde{B}\T P.
  \end{align*}
  Since $(\tilde{A},\tilde{B})$ is stabilizable and
  $\re(\lambda_i)>0$, it follows from LaSalle’s invariance principle
  that $\tilde{A}-\rho\lambda_i \tilde{B}\tilde{B}\T P$ is Hurwitz
  stable and we obtain the required stability of \eqref{cls3}.
	
  Meanwhile, from \cite[Lemma 2]{liu-zhang-saberi-stoorvogel-auto}, we
  can obtain the state synchronization result
  \[
    \lim_{t\to\infty}{z}_i- z_j\to0.
  \]
  Furthermore, it implies that
  \[
    \lim_{t\to\infty}{x}_i- x_j\to0.
  \]
  Therefore, the synchronization result can be obtained for any graph
  $\mathscr{G} \in \mathbb{G}^N$ with any size of the network $N$.
\end{proof}

\subsection{Protocol design for full-state coupling, i.e. $C=I$}


When $C=I$, we only need the assumption that the agents are
stabilizable and neutrally stable, i.e., the other conditions in
Assumption \ref{ass1} are satisfied automatically. 
Moreover, since
\eqref{zeta} can be rewritten as
\begin{equation}\label{zetax}
  \zeta_i= \sum_{j=1}^{N}\ell_{ij}x_j,
\end{equation}
it means that we do need neither a pre-compensator nor use SCB to
transform the compensated system \eqref{newsys}. Thus, we can obtain a static protocol, i.e., the estimator (or observer) is not needed to achieve the synchronization. 
Of course, the protocol design in \eqref{ncsfpvpscct}
can still be applied.

Firstly, since agent model \eqref{homo_sys} is neutrally stable, there
still exists a positive definite matrix $P$ such that
\begin{equation}
  PA+A\T P\leq 0. \label{codctf}
\end{equation}

The scale-free protocol design for continuous-time MAS with neutrally
stable agent is listed as follows.
\begin{tcolorbox}[breakable,colback=white]
  \begin{equation}\label{ncsfpvfscct}
    u_i=-\rho B\T P\zeta_i,
  \end{equation}
  where $P>0$ satisfies \eqref{codctf} and $\rho>0$.
\end{tcolorbox}

Then, we have the following theorem.
\begin{theorem}\label{th-full-linear}
  Consider a continuous-time MAS consisting of neutrally stable agents
  described by \eqref{homo_sys} and \eqref{zetax} where $(A,B)$ is
  stabilizable. Let the set $\mathbb{G}^N$ denote all graphs satisfy
  Definition \ref{def1}.
	
  Then, the scale-free state synchronization problem via linear protocol
  as stated in Problem \ref{prob4} is solvable. More specifically,
  then protocol \eqref{ncsfpvfscct} achieves state
  synchronization for any fixed graph $\mathscr{G} \in \mathbb{G}^N$ with
  any size of the network $N$.
\end{theorem}

\begin{proof}
  Combining \eqref{homo_sys} and \eqref{ncsfpvfscct}, we obtain
  \begin{equation}
    \dot{x}_i=Ax_i-\rho B\T P\sum_{j=1}^N \ell_{ij}x_j
  \end{equation}
  Then we have
  \begin{equation}\label{ccls3f}
    \dot{x}=(I\otimes A-\rho L\otimes (BB\T P))x
  \end{equation}
  by defining
  \[
    x=\begin{pmatrix}
      x_1\\\vdots\\x_N
    \end{pmatrix}.
  \]
  Similar to the proof of Theorem \ref{th-linear}, we can obtain the following transformed system
  \[
    \begin{system}{l}
      \dot{\phi}_1=A\phi_1, \\
      \dot{\phi}_i=(A-\rho \lambda_iBB\T P )\phi_i, \qquad i=2,\ldots,N
    \end{system}
  \]
  by using a non-singular matrix $T_f$. According to \cite[Lemma
  2]{liu-zhang-saberi-stoorvogel-auto}, we just prove the stability of
  $A-\rho \lambda_iBB\T P$ to obtain the state synchronization.
	
  Since $P>0$ satisfies \eqref{codctf}, we have
  \[
    P(A-\rho \lambda_iBB\T P)+(A-\rho \lambda_iBB\T P)^*P\leq -2\rho\re(\lambda_i)PBB\T P\leq 0
  \]
  for $\rho>0$.
  
  Since $(A,B)$ is stabilizable and $\re(\lambda_i)>0$, it follows
  from LaSalle’s invariance principle that $A-\rho \lambda_iBB\T P$ is
  Hurwitz stable. Thus, the synchronization result can be obtained for
  any graph $\mathscr{G} \in \mathbb{G}^N$ with any size of the
  network $N$.
\end{proof}

\section{Scale-free non-collaborative protocol design: Discrete-time case}

We make the following assumptions:
\begin{assumption}\label{ass2}
  Discrete-time agents \eqref{homo_sys} satisfy the following properties:
  \begin{enumerate}
  \item Agent model is stabilizable and detectable.
  \item Agent model is neutrally stable. 
  \end{enumerate}
\end{assumption}
\begin{remark}
  These assumptions are equal to  the necessary conditions we obtained for
  SISO systems in Theorem \ref{theo_nec2} 
\end{remark}

Meanwhile, there still exists a positive definite matrix $P$ such that
\begin{equation}
	A\T PA-P\leq 0.\label{coddtf}
\end{equation}

Our design is intrinsically different from the continuous-time. We
first start of with the partial-state coupling, which is going to use a stable observer with the so-called CSS architecture.
\subsection{Protocol design for partial-state coupling}



We have the following scale-free protocol design for discrete-time MAS
with neutrally stable agents:

\begin{tcolorbox}[breakable,colback=white]
	\begin{equation}\label{ncsfpvpscdt}
		\begin{system*}{ccl}
			\chi_i(t+1)&=&(A-HC) \chi_i(t)+H\zeta_i(t)   \\
			u_i(t)&=&-\delta B\T PA \chi_i(t) 
		\end{system*}
	\end{equation}
	where $\delta\in(0,\delta^*]$ and $\delta^*$ is obtained only from the knowledge of agent model ($A,B,C$). $P>0$ satisfies \eqref{coddtf} and
	$H$ is a matrix such that $A-HC$ is Schur stable. For computation of $\delta^*$, see the proof of Theorem \ref{th-linear-dt}.
\end{tcolorbox}

Then, we have the following theorem.
\begin{theorem}\label{th-linear-dt}
	Consider a discrete-time MAS described by \eqref{homo_sys} and
	\eqref{zeta-y}. Assume Assumption \ref{ass2} is satisfied. Let the
	set $\mathbb{G}^N$ denote all graphs satisfy Definition \ref{def1}.
	
	Then, the scale-free state synchronization problem via
	non-collaborative linear protocol as stated in Problem \ref{prob4}
	is solvable. More specifically, there exists $\delta^{*}>0$ which is obtained only from agent model ($A, B,C$), such that
	for all $\delta \in (0, \delta^*]$, the protocol
	\eqref{ncsfpvpscdt} achieves state
	synchronization for any fixed graph $\mathscr{G} \in \mathbb{G}^N$ with
	any size of the network $N$.
\end{theorem}

\begin{proof}
	For agent model \eqref{homo_sys} and \eqref{ncsfpvpscdt}, we have
	\begin{equation}\label{dcls2p}
		\begin{system}{ll}
			x(t+1)&=(I\otimes A)x(t)-[\delta(I-{D})\otimes BB\T PA]\chi(t),\\
			\chi(t+1)&=[I\otimes(A-HC)]\chi(t)+(I\otimes HC)x(t).
		\end{system}		
	\end{equation}
	By using \cite[Lemma 3]{liu-zhang-saberi-stoorvogel-ejc}, there
	exists a non-singular matrix $T_f$, we can transform \eqref{dcls2p}
	to
	\begin{equation}\label{dcls3p}
		\begin{system}{ll}
		\phi_i(t+1)&=A\phi_i(t)-\delta(1-\lambda_i) BB\T PA\psi_i(t), \\
		\psi_i(t+1)&=(A-HC)\psi_i(t)+HC\phi_i(t)
	   \end{system}
	\end{equation}
	for $i=2,\ldots, N$, where $\lambda_i$ satisfies $|\lambda_i|<1$.  Thus, we only need to prove that
	the system \eqref{dcls3p}
	is asymptotically stable for all $|\lambda_i|<1$.
	
	Define $e_i(t)=\phi_i(t)-\psi_i(t)$. The system \eqref{dcls3p} can be
	rewritten in terms of $\phi_i(t)$ and $e_i(t)$ as
	\begin{equation}\label{hodt-sys-cl3}
		\begin{system}{ccl}
			\phi_i(t+1) &=& (A-(1-\lambda_i)\delta BB\T PA)\phi_i(t)+(1-\lambda_i)\delta
			BB\T PA e_i(t),\\ 
			e_i(t+1) &=& (A-HC+(1-\lambda_i)\delta BB\T PA)e_i(t)\\
			&&\hspace{3.5cm}-(1-\lambda_i)\delta
			BB\T PA \phi_i(t). 
		\end{system}
	\end{equation}
	Let $Q$ be the positive definite solution of the Lyapunov equation,
	\[
	(A-HC)\T Q(A-HC)-Q+4I=0.
	\]
	There exists a $\delta_1$ such that for all
	$\delta\in(0,\delta_1]$, we have
	\begin{multline*}
		(A-HC+(1-\lambda_i)\delta BB\T PA)^* Q(A-HC+(1-\lambda_i)\delta BB\T PA)\\
		-Q+3I\leq 0. 
	\end{multline*}
	for all $\lambda_i$ with $|\lambda_i|<1$. Consider $V_1(t)=e_i(t)^*Qe_i(t)$ and let $\mu=\delta B\T PA \phi_i(t)$.  We have
	\begin{align*}
		&V_1(t+1)-V_1(t)  \\
		& \leq  -3\|e_i(t)\|^2+|1-\lambda_i|^2 \mu^{*}B\T QB \mu \\
		&\hspace*{1cm} + 2\left| \left((1-\lambda_i)^{*}
		\mu^{*}B\T Q[A-HC+(1-\lambda_i)\delta  BB\T PA]e_i(t)\right)\right| \\
		& \leq  -3\|e_i(t)\|^2+ |1-\lambda_i|^2
		M_{2}\|\mu\|^2 \\
		&\hspace*{1cm} + (|1-\lambda_i| M_{1}+ |1-\lambda_i|^2  \delta M_{3})\|\mu\|\|e_i(t)\|,
	\end{align*}
	where
	$
	M_1=2\|B\T Q\|\|A-HC\|$, $M_2=\|B\T QB\|
	$, and $M_3=2\|B\T Q\|\|BB\T PA\|$.
	It should be noted that $M_1$, $M_2$, and $M_3$ are independent of $\delta$
	and $\lambda$.	
	Consider $V_2(t)=\phi_i^{*}(t) P\phi_i(t)$. Note that
	\begin{multline*}
		[A-(1-\lambda_i)\delta BB\T PA]^* P[A-(1-\lambda_i)\delta BB\T PA]-P \\
		\leq -2\re(1-\lambda_i)\delta A\T PBB\T PA+|1-\lambda_i|^2\delta^2A\T PBB\T PBB\T PA.
	\end{multline*}
	There exists a $\delta_2<\delta_1$ such that, for all $\delta\in(0,\delta_2]$, we have
	$2\delta B\T PB\leq I_m$. Since
	$|1-\lambda_i|^2\leq 2\re(1-\lambda_i)$ for $|\lambda_i|<1$, we get for all
	$\delta\in(0,\delta_2]$,
	\begin{multline*}
		[A-(1-\lambda_i)\delta BB\T PA]^*P[A-(1-\lambda_i)\delta BB\T PA]-P \\
		\leq -\tfrac{1}{2}|1-\lambda_i|^2\delta A\T PBB\T PA. 
	\end{multline*}
	Hence 
	\begin{align*}
		V_2&(t+1)-V_2(t)\\
		  & \leq -\tfrac{1}{2\delta}|1-\lambda_i|^2 \|\mu\|^2+|1-\lambda|^2\delta^2e_i^*(t)A\T PBB\T PBB\T PAe_i(t)\\
		&\hspace{1cm} +2\left| (1-\lambda^*) e_i^*(t)A\T PB\mu-|1-\lambda|^2\delta
		e_i^*(t)A\T PBB\T PB\mu\right| \\
		& \leq  -\tfrac{1}{2\delta}|1-\lambda|^2\|\mu\|^2
		+\theta_1|1-\lambda|\|e_i(t)\|\|\mu\| \\
		&\hspace{3cm} +\theta_2\delta^2\|e_i(t)\|^2+\theta_3\delta|1 -\lambda|^2\|e_i(t)\|\|\mu\|,  
	\end{align*}
	where
	$
	\theta_1=2\|A\T PB\|$, $\theta_2=4\|A\T PBB\T PBB\T PA\|$, and $\theta_3=2\|A\T PBB\T PB\|$.	
	Define a Lyapunov candidate $V(t)=V_1(t)+\delta \kappa V_2(t)$ with
	$
	\kappa=4+2M_2+2M_1^2$.
	We get that
	\begin{align*}
		V(t&+1)-V(t)\\
		\leq&
		-(3-\delta^3\theta_2\kappa)\|e_i(t)\|^2-(2+M_1^2)|1-\lambda_i|^2\|\mu\|^2\\ 
		&\hspace{1cm}+(M_1+\delta\theta_1\kappa)|1-\lambda_i|\|\mu\|\|e_i(t)\| \\
		&\hspace{1cm}+(\delta M_3+\delta^2\theta_3\kappa)|1-\lambda_i|^2 \|\mu\|\|e(t)\|. 
	\end{align*}
	There exists a $\delta^*<\delta_2$ such that for a $\delta\in(0,\delta^*]$,
	$
	3-\delta^3\theta_2\kappa \geq 2.5$, $M_1+\delta\theta_1\kappa\leq
	2M_1$, and $\delta M_3+\delta^2\theta_3\kappa\leq 1$.
	This yields,
	\begin{align*}
		V(t&+1)-V(t)   \\
		 \leq& -2.5\|e_i(t)\|^2 -(2+M_1^2)|1-\lambda_i|^2\|\mu\|^2\\
		& \hspace*{2cm} +(2M_1|1-\lambda_i| +|1-\lambda_i|^2)\|\mu\|\|e_i(t)\|\\
		 \leq& -0.5\|e_i(t)\|^2 -|1-\lambda_i|^2\|\mu\|^2 -
		(\|e_i(t)\|-M_1|1-\lambda_i|\|\mu\|)^2\\
		& \hspace*{2cm} -|1-\lambda_i|^2(\tfrac{1}{2}\|e_i(t)\|-\|\mu\|)^2\\
		\leq&  -0.5\|e_i(t)\|^2 -|1-\lambda_i|^2\|\mu\|^2.
	\end{align*}
	Since $(A,B)$ is controllable, it follows from LaSalle's invariance
	principle that the system \eqref{hodt-sys-cl3} is globally
	asymptotically stable for $\delta\leq\delta^{*}$.
\end{proof}
\subsection{Protocol design for full-state coupling, i.e. $C=I$}


Firstly, the information measurement \eqref{zeta-y} is rewritten as 
\begin{equation}\label{zeta-dx}
	\zeta_i(t)=\sum_{j=1,j\neq i}^N d_{ij}(x_i(t)-x_j(t)).
\end{equation}
The scalable protocol for discrete-time MAS with
neutrally stable agent via full-state coupling is listed as follows.
\begin{tcolorbox}[breakable,colback=white]
  \begin{equation}\label{ncsfpvfscdt}
    u_i(t)=-\eps B\T PA\zeta_i(t),
  \end{equation}
  where $P>0$ satisfies \eqref{coddtf}, and $\eps\in (0, \eps^*]$ with $\eps^*=\|B\T PB\|^{-1}$.
\end{tcolorbox}

Then, we have the following theorem.
\begin{theorem}\label{th-full-linear-dt}
  Consider a discrete-time MAS consisting of neutrally stable agents
  described by \eqref{homo_sys} and \eqref{zeta-dx} where $(A,B)$ is
  stabilizable. Let the set $\mathbb{G}^N$ denote all graphs satisfy
  Definition \ref{def1}.
	
  Then, the scale-free state synchronization problem via linear protocol
  as stated in Problem \ref{prob4} is solvable. More specifically, for
  any given $\eps\in (0, \eps^*]$ with $\eps^*=\|B\T PB\|^{-1}$, the
  protocol \eqref{ncsfpvfscdt} achieves state synchronization
  for any fixed graph $\mathscr{G} \in \mathbb{G}^N$ with any size of the
  network $N$.
\end{theorem}

\begin{proof}	
  For agent model \eqref{homo_sys} and \eqref{ncsfpvfscdt}, we have
  \begin{equation}\label{dcls2f}
    x(t+1)=[I\otimes A-\eps(I-{D})\otimes BB\T PA]x(t).
  \end{equation}
  By using \cite[Lemma 3]{liu-zhang-saberi-stoorvogel-ejc}, there
  exists a non-singular matrix $T_f$, we can transform \eqref{dcls2f}
  to
  \begin{equation}\label{dcls3f}
    \phi_i(t+1)=(A-\eps(1-\lambda_i) BB\T PA)\phi_i(t), \quad i=2,\ldots, N,
  \end{equation}
  where $\lambda_i$ satisfies $|\lambda_i|<1$.  Thus, we just need to
  prove the stability of $A-\eps(1-\lambda_i) BB\T PA$.
	
  Since the
  matrix $P>0$, we obtain the stability of \eqref{dcls3f}.
  \begin{equation*}
    [A-\eps(1-\lambda_i) BB\T PA]^*P[A-\eps(1-\lambda_i) BB\T PA]-P
    \leq -\varphi A\T PBB\T PA
  \end{equation*}
  with $\varphi= \eps [ 2\re(1-\lambda_i)-|1-\lambda_i|^2]$. Note
  that $|\lambda_i|<1$ implies
  \begin{equation}\label{dfbound}
    |1-\lambda_i|^2\leq 2\re(1-\lambda_i),
  \end{equation}
  and therefore we have $\varphi>0$. Since $(A,B)$ is stabilizable, it then follows from
  LaSalle’s invariance principle that the system \eqref{dcls3f} is
  globally asymptotically stable. Note that $\eps^*$ depends only on
  agent’s model, hence the synchronization result can be obtained for
  any graph $\mathscr{G} \in \mathbb{G}^N$ with any size of the
  network $N$.
\end{proof}
\begin{remark}
	The results in \cite{liu-zhang-saberi-stoorvogel-auto} and \cite{liu-zhang-saberi-stoorvogel-ejc} are used in Theorems \ref{th-linear}-\ref{th-full-linear-dt}. Compared with this paper, \cite{liu-zhang-saberi-stoorvogel-auto} focused on continuous-time MAS with agents which are squared-down passive and passifiable. The linear protocol for squared-down passive agents is scalable and a subset of the design in this paper. In particular, the additional structure in [11] enabled the use of static protocols which is not possible for the more general class of agents in this paper. The nonlinear adaptive protocols are also scalable for the undirected communication network.	
	\cite{liu-zhang-saberi-stoorvogel-ejc} developed a linear protocol design for discrete-time MAS
	only with squared-down passifiable via input feedforward agents. The
	designs is not scale-free.
\end{remark}

\section{Numerical Examples}

In this section, we will illustrate the effectiveness of our designs
with two numerical examples for state synchronization of continuous-
and discrete-time MAS with partial-state coupling. Meanwhile, we
consider two communication networks with different topologies to show
the scalability of our protocols.

\begin{figure}[h]
	\includegraphics[width=6.5cm, height=1.cm]{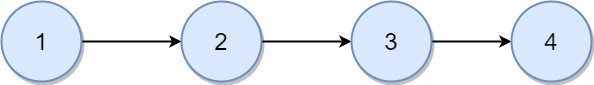}
	\centering
	\caption{Directed topology network with $4$ nodes}\label{graph1}
\end{figure}

\textbf{Case $I$:} We consider MAS with $4$ agents $N =4$, and directed communication
topology shown in Figure \ref{graph1}. 

\textbf{Case $II$:} In this case, we consider MAS with $60$ agents
i.e. $N = 60$, and directed communication topology with associated adjacency matrix $\mathcal{A}_{II}$ being $a_{i+1,i}=a_{1,60}=1$ and $i=1,\cdots,59$. 

Then, the continuous- and discrete-time MAS are studied respectively.

\subsubsection{Continuous-time MAS}
Consider continuous-time agent models \eqref{homo_sys} with the
following parameter:
\begin{equation*}
  A=\begin{pmatrix}
    0&1&1\\-1&0&1\\0&0&0
  \end{pmatrix},\  B=I,\ C=\begin{pmatrix}
    1&0&0\\
    0&1&0
  \end{pmatrix}.
\end{equation*}
We design pre-compensator \eqref{precomz} with the choice of 
\[
  A_p=-2,\ B_p=1,\ C_p=\begin{pmatrix} 0&0&1
\end{pmatrix}\T,\ D_p=\begin{pmatrix}
	0&1&0
\end{pmatrix}\T.
\]
Then the other protocol parameters in \eqref{ncsfpvpscct} are as follows,
\begin{align*}
  P=\begin{pmatrix}
    1&0 &-1 &-0.6\\
    0& 1& 1& 0.2\\
    -1& 1& 3& 1.3\\
    -0.6& 0.2& 1.3& 2
  \end{pmatrix},
  S^{-1}=\begin{pmatrix}
    1&0 &0 &0\\
    0& 0& 0& 1\\
    0& 0& 1&0\\
    0& -1& 0& 1
  \end{pmatrix}, \tilde{B}=\begin{pmatrix}
    0\\
    1\\
    0\\
    1
  \end{pmatrix},\\
  A_{11}=\begin{pmatrix}
    0& 0 &1\\
    -1& -2& 1\\
    0& -1& 0
  \end{pmatrix},
  A_{12}=\begin{pmatrix}
    1\\2\\1
  \end{pmatrix}, H=\begin{pmatrix}
    1\\0\\1
  \end{pmatrix}, \bar{C}=\begin{pmatrix}
    1&0&0
  \end{pmatrix}.
\end{align*}

\begin{figure}[h!]
	\includegraphics[width=9.5cm]{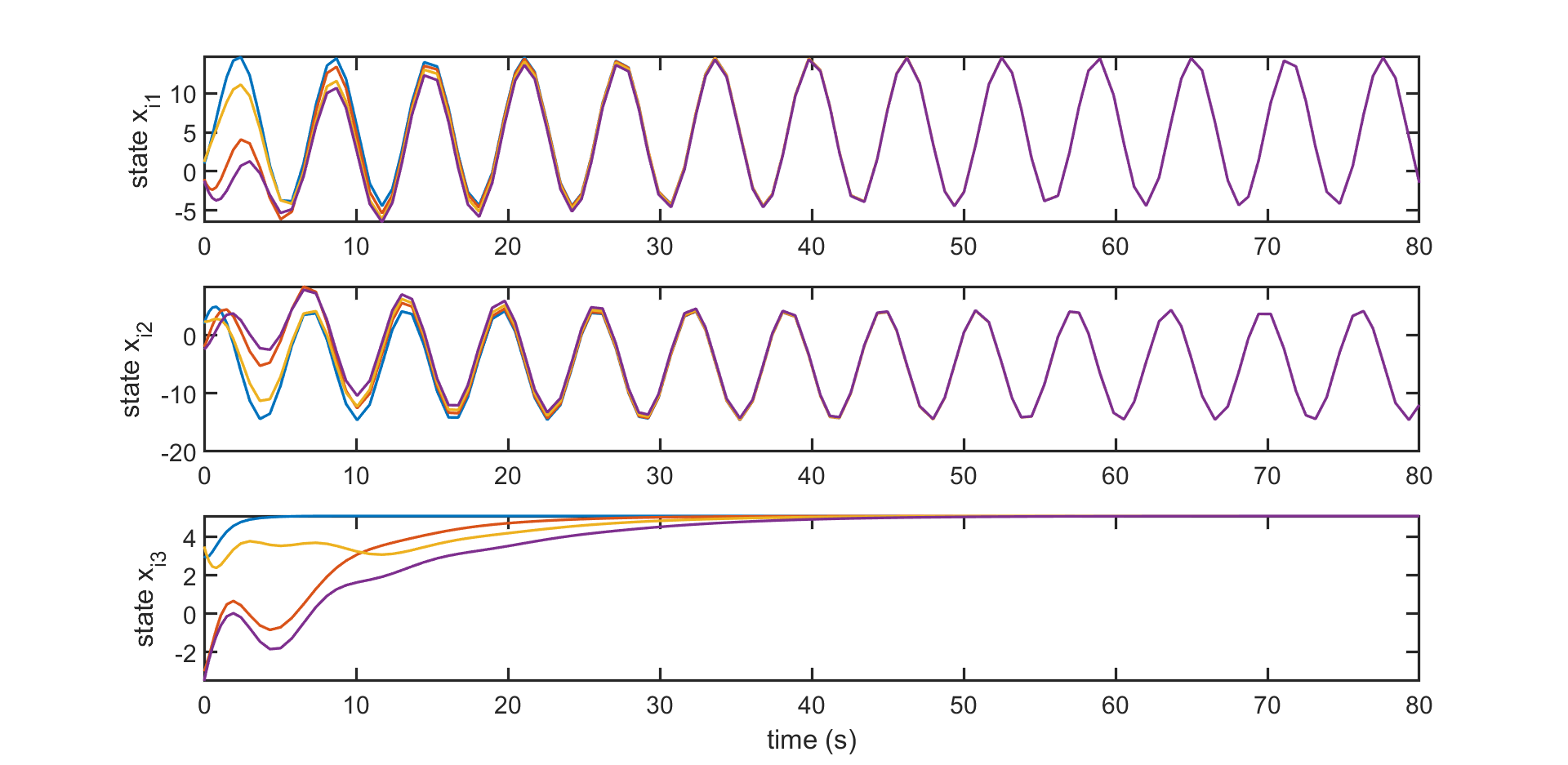}
	\centering
	\caption{State synchronization for continuous-time MAS with communication graph in Case $I$.}\label{con-case1}
\end{figure}

\begin{figure}[h!]
  \includegraphics[width=9.5cm]{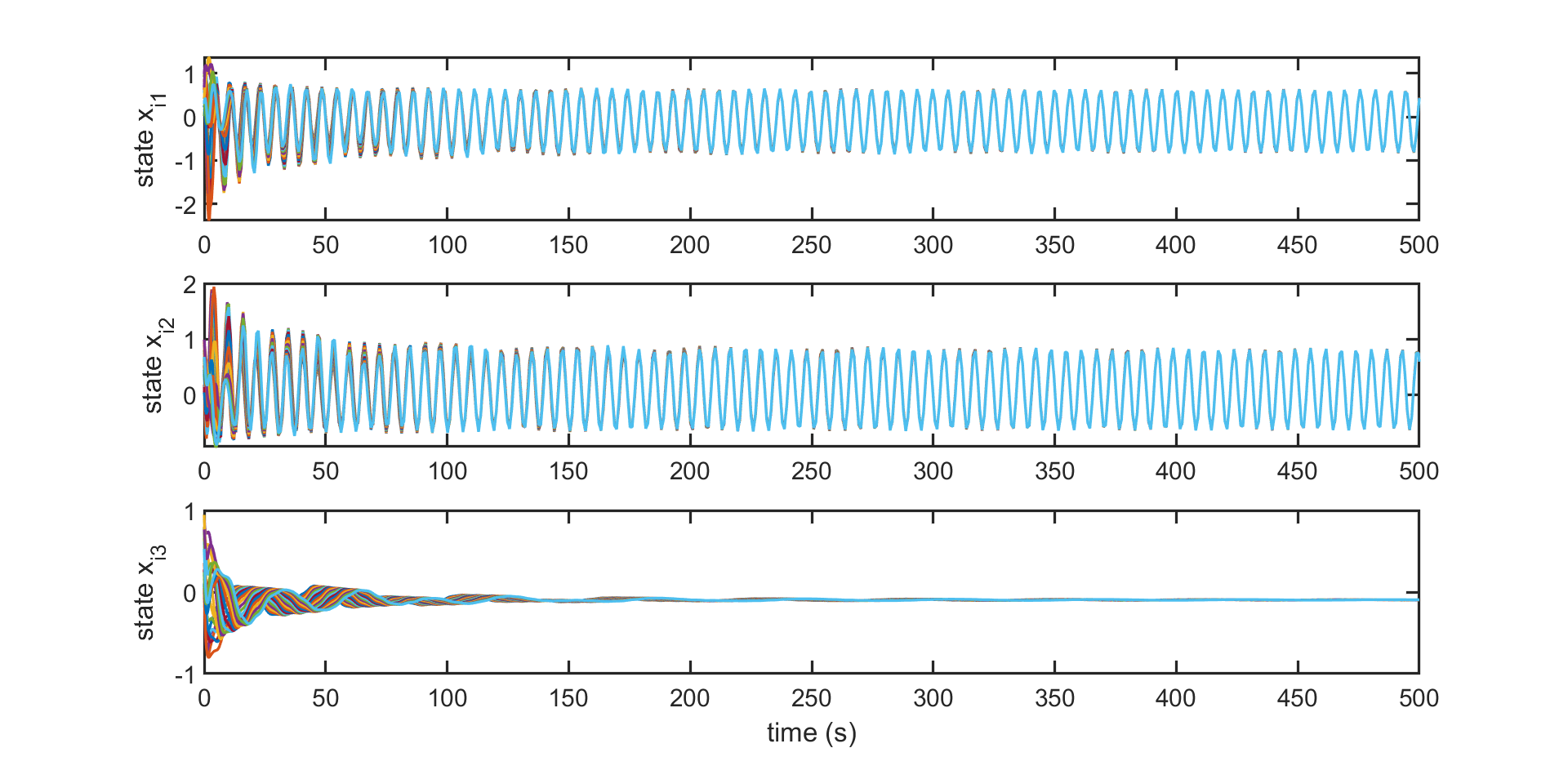} \centering
  \caption{State synchronization for continuous-time MAS with communication graph in Case $II$.}\label{con-case2}
	\includegraphics[width=9.5cm]{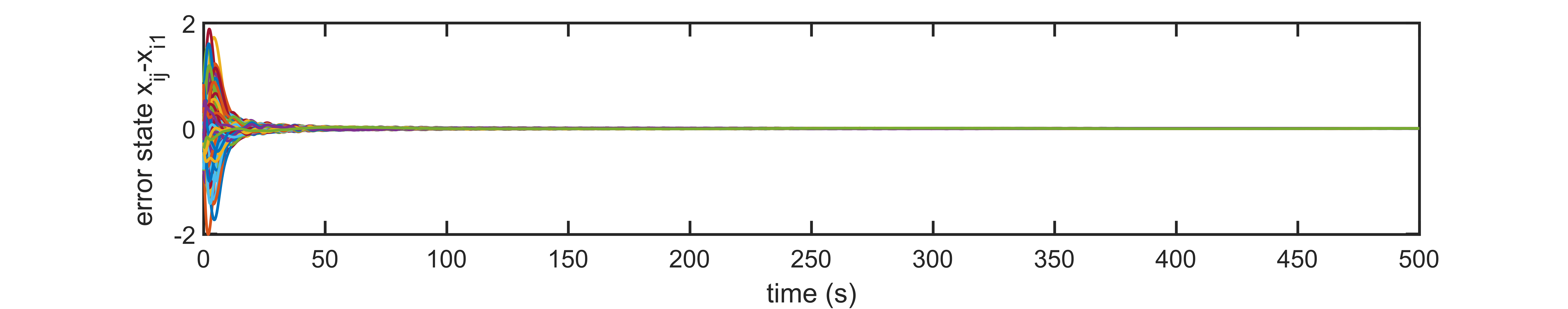} \centering
	\caption{Error state for continuous-time MAS with communication graph in Case $II$.}\label{con-case2-e}
\end{figure}

The simulation results for both Cases I and II are demonstrated in
Figure \ref{con-case1} and \ref{con-case2}. And the error states $x_{ij}-x_{i1}$ are shown in \ref{con-case2-e} to show the synchronization more clearly. The results show that the
protocol design is independent of the communication graph and is scale
free so that we can achieve synchronization with one-shot protocol
design, for any graph with any number of agents.


Compared with scale-free collaborative protocol design in \cite{liu-saberi-stoorvogel-donya-almost-automatica}, the synchronized time is deteriorating since no extra information exchange is employed. For example, the running time of the 60-node case is 21.6768s, but this time is 6.8072s under the same parameters using the scalable collaborative protocol by SIMULINK. However, non-collaborative protocol design does not need extra information exchange through communication network and is more likely applied in practical.\medskip

\subsubsection{Discrete-time MAS}

Consider discrete-time agent models \eqref{homo_sys} with the
following parameters:
\begin{equation*}
  A=\begin{pmatrix}
    0&1&1\\-1&0&1\\0&0&1
  \end{pmatrix},\ B=\begin{pmatrix}
  0\\0\\1
\end{pmatrix},\ C=\begin{pmatrix}
    1&0&0
  \end{pmatrix}.
\end{equation*}

We design protocol \eqref{ncsfpvpscdt}  with the following parameters 
\[
H=\begin{pmatrix} 0.5\\ -0.5 \\0.4
\end{pmatrix},\ P=\begin{pmatrix}
1&0 &-1\\
0& 1& 0\\
-1& 0& 2
\end{pmatrix}, \delta=0.1.
\]

\begin{figure}[h!]
  \includegraphics[width=9.5cm]{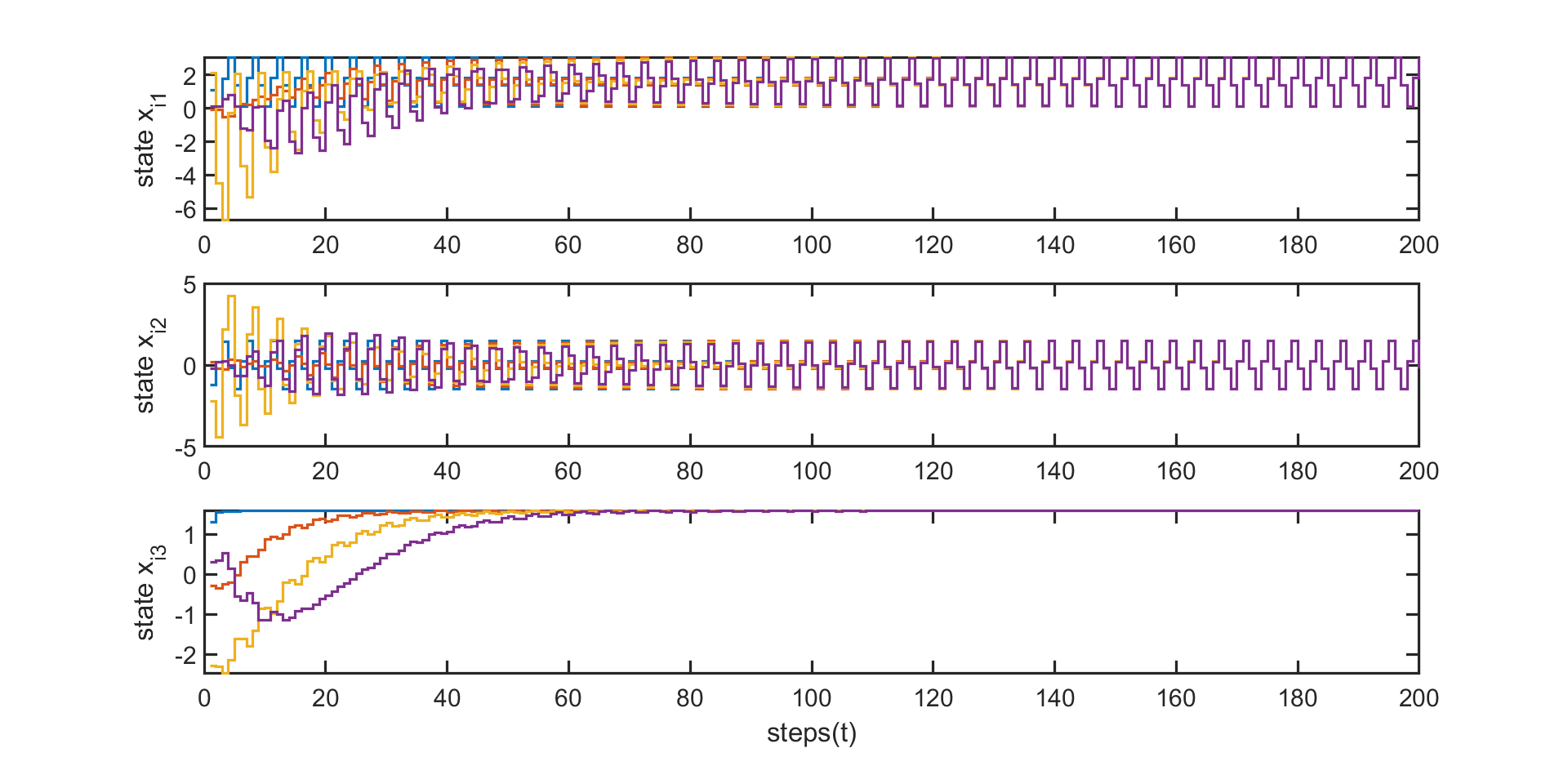}
  \centering
  \caption{State synchronization for discrete-time MAS with communication graph in Case $I$.}\label{dis-case1}
\end{figure}

\begin{figure}[h!]
  \includegraphics[width=9.5cm]{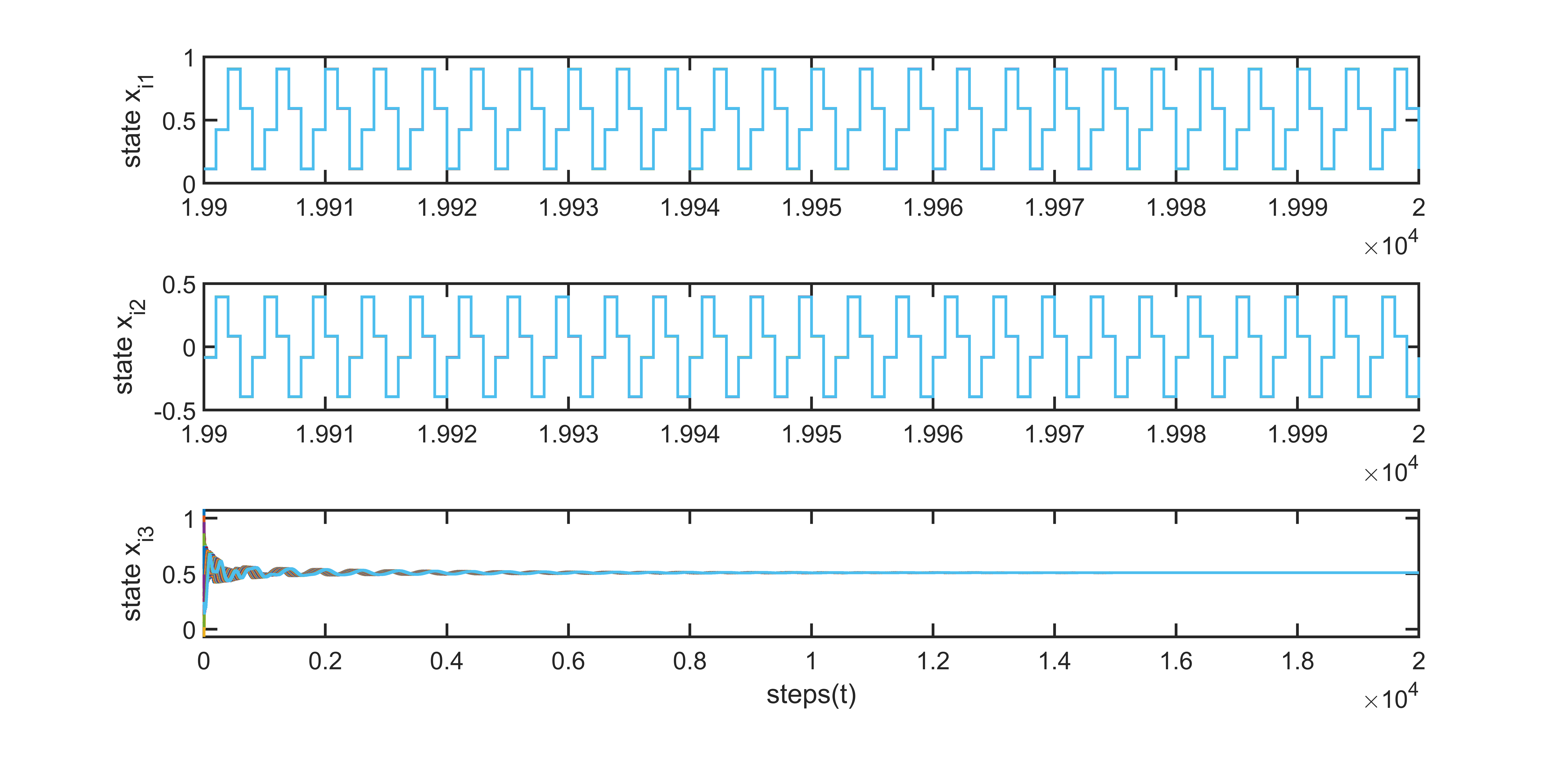}
  \centering
  \caption{State synchronization for discrete-time MAS with communication graph in Case $II$. In particular, we only show the synchronized trajectories for states $x_{i1}$ and $x_{i2}$, since there are many lines to make the figures difficult illuminating.}\label{dis-case2}
  \vspace{0.3cm}
    \includegraphics[width=9.5cm]{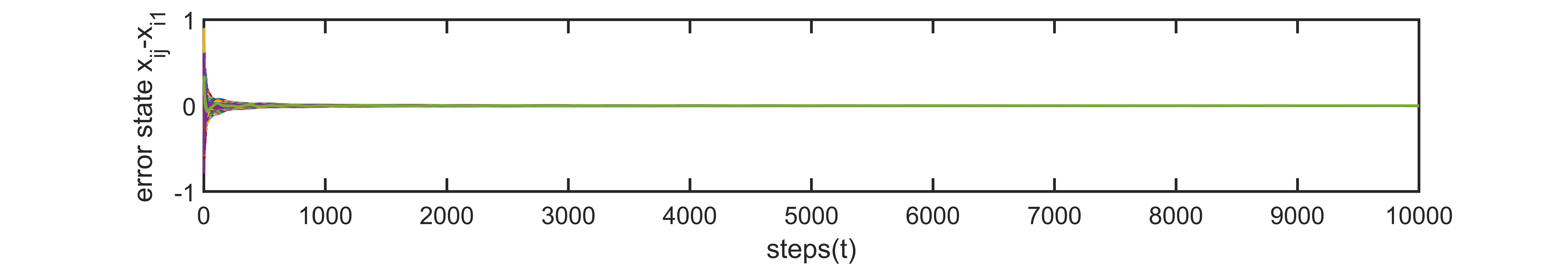}
  \centering
  \caption{Error state for discrete-time MAS with communication graph in Case $II$.}\label{dis-case2-e}
\end{figure}

We let the information exchange $\zeta_{i}$ satisfy \eqref{zetabar2-d0}. Then the simulation results for both Cases I and II are demonstrated in Figure \ref{dis-case1} and \ref{dis-case2}. And the error states $x_{ij}-x_{i1}$ are shown in \ref{con-case2-e} to show the synchronization more clearly. The results show that the protocol design is independent of the communication graph and is scale free so that we can achieve synchronization with one-shot protocol design, for any graph with any number of agents.

\section{Conclusion}
In this paper, we have proposed a scale-free non-collaborative
protocol design to achieve state synchronization for homogeneous MAS
with the agents satisfying Assumptions \ref{ass1} and \ref{ass2}. Moreover, we have provided these assumptions (conditions) are very close to necessary. The
non-collaborative protocols are designed for one class of continuous- and discrete-time	
MAS, which are solely based on agent models
without utilizing localized collaborative information
exchange, and work for any number of agents and any fixed communication graph
containing a spanning tree.

%
%
%

\bibliographystyle{plain}
\bibliography{referenc}

\end{document}